\documentclass[pre,twocolumn,showpacs,preprintnumbers,amsmath,amssymb]{revtex4}

\usepackage{graphicx}
\usepackage{bm}
\usepackage{hyperref}
\hypersetup{colorlinks = true, urlcolor = green, linkcolor = green, citecolor = green}
\usepackage{color}
\usepackage{ulem}

\newcommand{\AT}[1]{\textcolor{black}{#1}}

\begin{document}

\title{Propagative and diffusive regimes of acoustic damping in bulk amorphous material}

\author{Y. M. Beltukov}
\affiliation{Ioffe Institute, 194021 St. Petersburg, Russian Federation}
\affiliation{Universit\'e Montpellier II, CNRS, Montpellier 34095, France}
\author{D. A. Parshin}
\affiliation{Peter the Great St. Petersburg Polytechnic University, 195251 St. Petersburg, Russian Federation}
\author{V. Giordano}
\affiliation{Universit\'e de Lyon, LaMCoS, INSA-Lyon, CNRS UMR5259, F-69621, France}
\affiliation{Institut Lumi\`ere  Mati\`ere, UMR 5306 Universit\'e Lyon 1-CNRS,  F-69622 Villeurbanne Cedex, France}
\author{A. Tanguy}
\affiliation{LaMCos, INSA-Lyon, CNRS UMR5259, Universit\'e de Lyon, F-69621 Villeurbanne Cedex, France}

\begin{abstract}
    In amorphous solids, a non-negligible part of thermal conductivity results from phonon scattering on the structural disorder. The conversion of acoustic energy into thermal energy is often measured by the Dynamical Structure Factor (DSF) thanks to inelastic neutron or X-Ray scattering. The DSF is used to quantify the dispersion relation of phonons, together with their damping.  However, the connection of the dynamical structure factor with dynamical attenuation of wave packets in glasses is  still a matter of debate. We focus here on the analysis of wave packets propagation in numerical models of amorphous silicon. We show that the DHO fits (Damped Harmonic Oscillator model) of the dynamical structure factors give a good estimate of the wave packets mean-free path, only below the Ioffe-Regel limit. Above the Ioffe-Regel limit and below the mobility edge, a pure diffusive regime without a definite mean free path is observed. The high-frequency mobility edge is characteristic of a transition to localized vibrations. Below the Ioffe-Regel criterion, a mixed regime is evidenced at intermediate frequencies, with a coexistence of propagative and diffusive wave fronts. The transition between these different regimes is analyzed in details and reveals a complex dynamics for energy transportation, thus raising the question of the correct modeling of thermal transport in amorphous materials.
\end{abstract}

\pacs{%
61.43.Dq, 
61.43.Fs, 
63.50.-x, 
65.60.+a, 
62.20.-x  
}

\maketitle

\section{Introduction}
\label{sec:intro}

It is today largely known that amorphous materials, despite the lack of periodicity, can sustain collective vibrations remniscent of phonons in crystals. Such vibrations have however glassy-specific features, that arise when their wavelength becomes comparable to the disorder lengthscale, which happens at frequencies of few THz. Anomalous vibrations have so been observed by means of inelastic X-Ray or Neutron scattering~\cite{Boon1980, giordano_2011,Chen2015}, or Raman spectroscopy~\cite{Duval2007,Carini2015,Tanguy2015}. One specific property of amorphous materials is an anomalous density of vibrational states at frequencies of a few THz, the so-called ``Boson Peak'': an excess of modes with respect to the Debye prediction. The structural origin of the Boson Peak was extensively discussed~\cite{Schirmacher1998, Taraskin2001, Leonforte2006, Parshin2007, Monaco2008, Chumakov2011}, especially because it enhances the thermal capacity in glasses~\cite{Zeller1971}, but the nature of the vibrations composing the Boson peak, was never clearly identified~\cite{Taraskin2001, Schirmacher2007, Ruffle2008, Chumakov2011}, nor was their diffusive or propagative" character": it means,  are they related to diffusive or propagative transportation of an external excitation?
Such modes have been related to the low value of the thermal conductivity in glasses and more specifically to the characteristic plateau appearing in its temperature dependence at low temperature ($\approx$ 10~K)~\cite{Zeller1971}. While the thermal conductivity is easily computed in crystals, it is far more difficult to compute it in glasses, due to the lack of periodicity and the intricate nature of phonons in this case, as well as the unknown attenuation mechanisms at play in these materials~\cite{Feldman1993,b.tanguy2002,Wyart2010, Beltukov2013}. Recently, it has been shown that the thermal diffusivity in amorphous materials is controlled by scattering processes~\cite{taraskin2002,Page2009,Beltukov2013,larkin,Beltukov2016,Lemaitre2016}. Three regimes were first evidenced: a plane-wave  dominated regime with a low scattering of transverse and longitudinal waves, then a regime of strong scattering above the Ioffe-Regel criterion, and finally Anderson localization near the mobility edge~\cite{Beltukov2016}. In the present paper, we focus on an already well-studied~\cite{b.stillinger,b.allen,larkin} model amorphous material ($\nu$-Si) to investigate the different modes of energy transfer for acoustic excitations at different frequencies. We compare explicitly the mean-free path inferred from the numerical measurement of the Dynamical Structure Factor and the attenuation length of the energy in a vibrational wave packet excitation, and we study in details the long-term dynamics of the materials in response to an acoustic excitation.

The paper is organized as follows: in the next section we present the numerical model of the material studied, and the efficient numerical method used to simultaneously compute wave packets propagation at all frequencies. Then we detail the different regimes of energy transfer as a function of the frequency. A special focus is done on the transition between the propagative and the diffusive regime. The origin of the attenuation of the maximum energy in the wave packets is discussed in connection to coherent and incoherent excitation. The results are finally compared to the analysis of the Dynamical Structure Factor.

\section{Numerical model}

\subsection{Model material}

We have studied the vibrational properties of a model amorphous silicon (a-Si) system consisting of $N=262\,144$ atoms contained in a periodic cubic box of lengths $L_x=L_y=L_z \approx 174$ \AA{}, that are replicated three times in $x$ direction. The technical details of the preparation of the a-Si model have already been presented in Ref.~\cite{Beltukov2016, b.fusco2010}. The atomic configurations of $\nu$-Si structure have been obtained using the open source LAMMPS package~\cite{b.lammps} for classical Molecular Dynamics simulations. The system is first prepared in a cubic diamond crystalline state, then heated and equilibrated in the liquid state, before rapid quench. The interactions are tuned along the quenching protocol in order to ensure a realistic percentage of coordination defects at rest, as detailed in~\cite{b.fusco2010}.
The Si-Si interaction in the system studied at rest is described by the empirical Stillinger-Weber potential~\cite{b.stillinger}, including two-body and three-body directional interactions to account for the covalent nature of the bonds. After the quenching, the system obtained is fully amorphous, without any intrinsic structural lengthscale despite an interatomic distance of $2.34$ {\AA}.  In a previous paper, we have studied in details the role of the three-body interactions on the vibrational density of states, on the phonons mean-free path and on the diffusivity in such systems~\cite{Beltukov2016}. The mean-free path was defined from the inverse phonon lifetime measured from the Dynamical Structure Factor (DSF) and the phonons group velocity, while the diffusivity was measured from the diffusive energy transfer resulting from the incoherent excitation of a central layer with a quasi-monochromatic wave packet. One open question concerns the relation between the mean free path measured from the DSF and the attenuation length of a wave packet generated from a coherent excitation. This is the focus of the present paper. Coherent and incoherent wave packets excitations (corresponding to ultrasonic or thermal excitations respectively) will also be compared.

\begin{figure*}[t!]
    \includegraphics[scale=0.55]{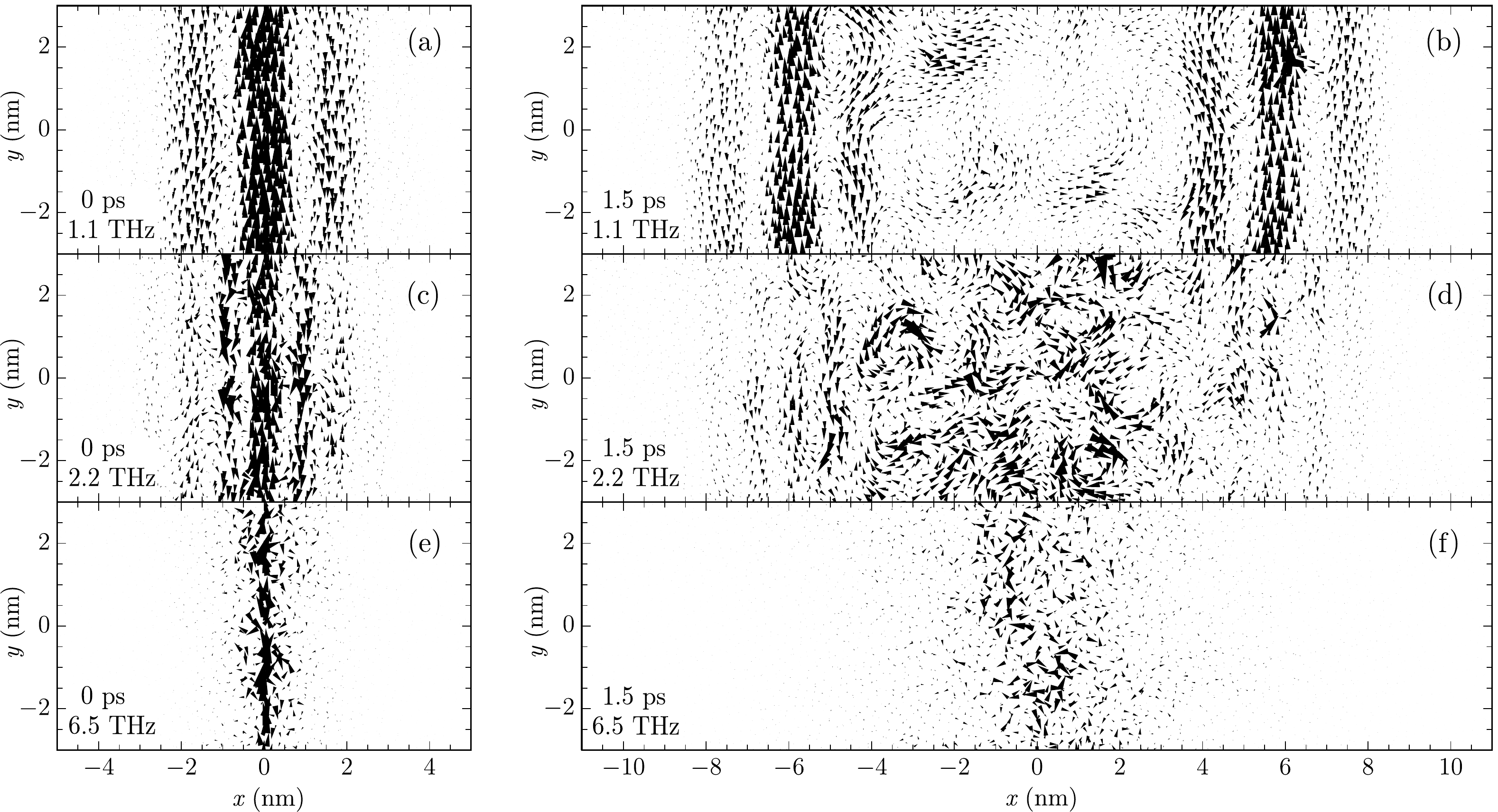}
    \caption{Atomic velocities during the action of the transverse excitation force (left column) and 1.5 ps after it (right column). The length and the direction of the arrow show the atomic velocity in $xy$ plane. Atoms within 0.5 nm slice in $z$ direction are shown. The frequency of the external force is shown on each panel. The real part of complex velocities is shown here.}
    \label{f.velocities}
\end{figure*}

\begin{figure}[t]
   \includegraphics[scale=0.55]{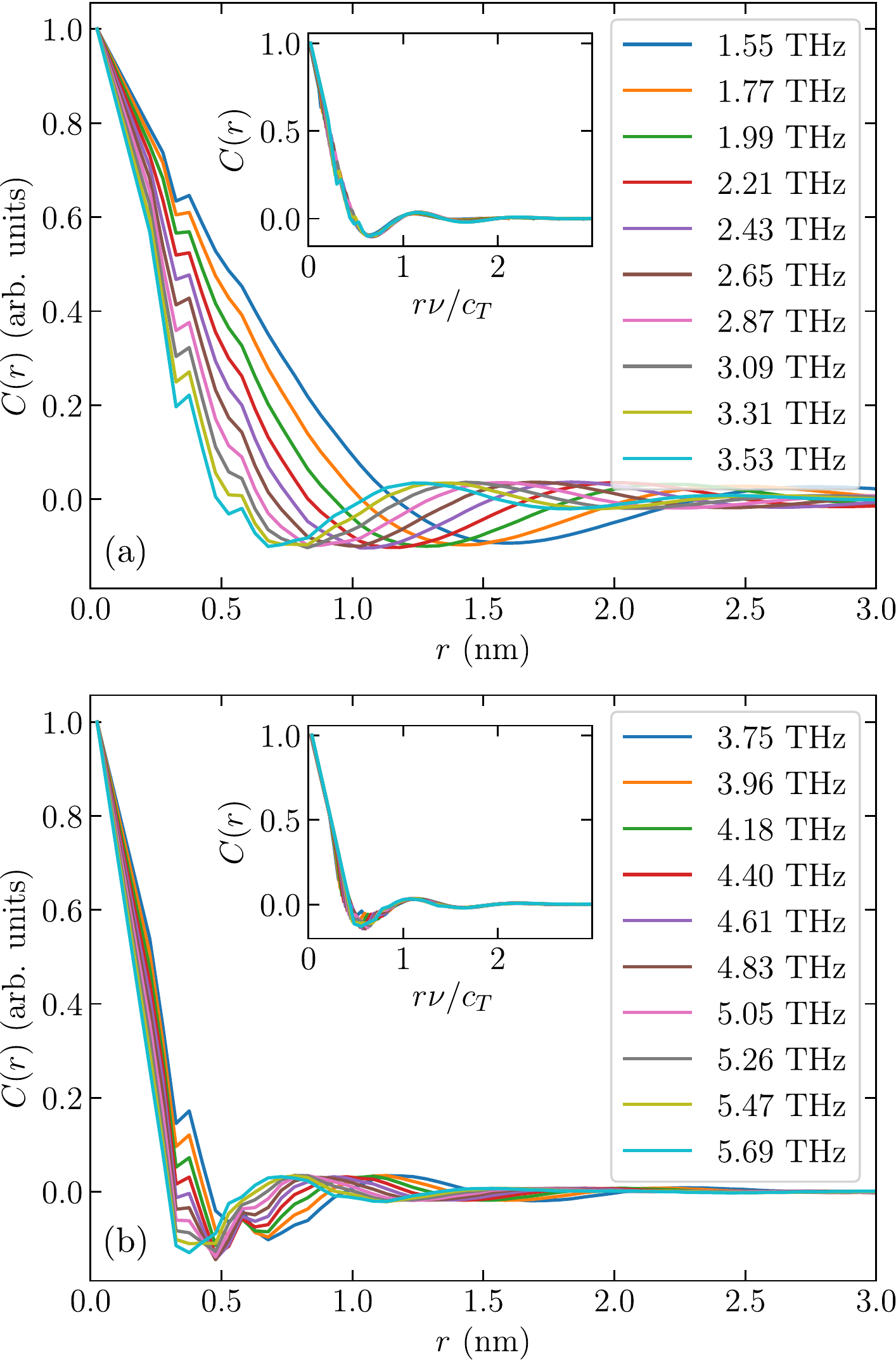}
   \caption{(Color online)  Spatial correlation function of the displacements excited behind the transverse wave front ($-3$ nm $ < x <  3$ nm) for $t = 1.5$ ps as a function of the distance $r$. Insets show the correlation function as a function of a rescaled distance $r\nu/c_T$.}
    \label{f.corr}
\end{figure}

\subsection{Dynamical Structure Factor}

The dynamical structure factor is the self-correlation function of the mass currents~\cite{shintani} in the system at thermal equilibrium at a temperature $T$. At a small enough temperature the structure factor can be calculated by normal mode analysis as detailed in~\cite{Beltukov2016} for the same system as the one studied in this paper.
In order to extract information on phonons in the low-frequency region, the structure factor $S_\eta(q,\omega)$ can be fitted using a DHO (Damped Harmonic Oscillator)  model.
\begin{equation}
    S_\eta(q,\omega) = \frac{A}{(\omega^2-\omega_\eta^2(q))^2+\omega^2\Gamma^2}, \quad \eta=L,T.   \label{eq:fit}
\end{equation}
where $\eta$ denotes longitudinal (L) or transverse (T) component.
The phonon dispersion $\omega_\eta(q)$ and inverse lifetime $\Gamma(q)$ can thus be obtained. They will be successively compared to the dynamical characteristics extracted from the wave packets propagation.

\subsection{Excitation of wave packets}

In order to study the wave packets propagation, we apply a quasi-monochromatic external pulse to a thin atomic layer around $x=0$. 
The excitation is applied along a cross-section with surface $L_yL_z$.
Only the harmonic response is studied here. In this case, the equation of motion has the form
\begin{equation}
    \ddot{u}_{i\alpha}(\omega, t) + \sum_{j\beta} M_{i\alpha,j\beta} u_{j\beta}(\omega, t) = f_{i\alpha}^{\rm exc}(\omega, t),   \label{eq:Newton_exc}
\end{equation}
where ${\bf u}_i=\sqrt{m}({\bf r}_i - {\bf R}_i)$ is a scaled displacement of the $i$-th atom from the equilibrium position ${\bf R}_i$ and $M_{i\alpha,j\beta}$ is an element of the dynamical matrix $M$. The excitation force $f_{i\alpha}^{\rm exc}(\omega, t)$ is the $\alpha$ component of a complex excitation force
\begin{equation}
    {\bf f}_i^{\rm exc}(\omega, t) = {\bf f}_\eta \exp\left(i\omega t-\frac{t^2}{2\tau_{\rm exc}^2}-\frac{x_i^2}{2w^2}\right).   \label{eq:exc}
\end{equation}

The width of the excited layer is $w=1$~{\AA} and the duration of the excitation is $\tau_{\rm exc} = 0.36$~ps. Such pulse duration is smaller than a typical phonon lifetime and gives a good enough frequency resolution $\delta\nu \sim 1/(2\pi\tau_{\rm exc})=0.4$~THz. The direction of the applied force is defined by ${\bf f}_\eta$, which is common for all atoms in the excited layer when the excitation is {\it coherent}, and which is randomly uniformly distributed in case of {\it incoherent} excitation. The subscript $\eta$ indicates the wave-packet polarization in case of coherent excitation. In this case, we use ${\bf f}_L$ in $x$ direction for the longitudinal polarization and ${\bf f}_T$ in $y$ direction for the transverse one. 

In order to study the wave packets propagation for different frequencies $\omega$ and both polarizations, we calculate the kinetic energy density averaged over the cross-section $L_yL_z$
\begin{equation}
    E_\eta(\omega, x, t) = \frac{1}{2L_yL_z}\sum_i |\dot{\bf u}_i^\eta(\omega, t)|^2\delta(x - x_i).
\end{equation}

We consider the kinetic energy only because it can be prescribed to atomic locations in a simple and unique way in contrast to the potential energy. The kinetic energy continuously transforms to the potential energy and backward with a frequency $2\omega$. In order to suppress these oscillations we applied the complex excitation force ${\bf f}_i^{\rm exc}(\omega,t)$ and used the notion of complex velocities $\dot{\bf u}_i^\eta(\omega, t)$. The real and imaginary parts of ${\bf f}_i^{\rm exc}(\omega,t)$ and $\dot{\bf u}_i^\eta(\omega, t)$ have a natural phase shift $\pi/2$. These parts are summed up in the definition of $E_\eta(\omega, x, t)$. In order to improve the efficiency of the numerical calculation it is also possible to use an impulse $\delta(t)$ and to compute the response to all the frequencies simultaneously. The method used in this work is detailed in Appendix~\ref{sec:WP}.

\section{Description of Wave packets propagation}


Fig.~\ref{f.velocities} shows the real part of atomic velocities during the action of the excitation force at different frequencies (left column) and 1.5 ps after the excitation (right column). We present the result for the transverse polarization and three different frequencies: 1.1 THz (top row -- a, b), 2.2 THz (middle row -- c, d) and 6.5 THz (bottom row -- e, f). Since the Ioffe-Regel frequency for transverse polarization is approximately $\nu_{\rm IR}^T \approx 4$ THz in this sample~\cite{Beltukov2016}, the frequencies 1.1 THz and 2.2 THz are definitely below $\nu_{\rm IR}^T$, whereas the frequency 6.5 THz is definitely above $\nu_{\rm IR}^T$.

In Fig.~\ref{f.velocities}(a) one can see the initial transverse wave packet, excited by the external force with a frequency 1.1 THz. This wave packet has a well recognizable plane-wave structure. After 1.5 ps (Fig.~\ref{f.velocities}(b)) we can see two separate wave packets, which move in opposite directions. The background between these wave packets (for $-3$~nm~$ < x < 3$~nm in the figure) originates from the low amplitude scattering of the wave packets during their propagation on the amorphous structure.

The wave packet in Fig.~\ref{f.velocities}(c) is similar to Fig.~\ref{f.velocities}(a) but for a frequency 2.2 THz: one can see a plane-wave structure with a smaller wavelength. However, after 1.5 ps (Fig.~\ref{f.velocities}(d)) the plane-wave structure is mixed with a random rotational motion with similar amplitude, left behind the front. In this case, as will be shown in the next part, the amplitude of the propagative planar excitation decreases with time and is progressively exceeded by the rotational one. In order to identify a correlation radius in the vortex structures shown in Fig.~\ref{f.velocities}, we have plotted in Fig.~\ref{f.corr} the correlation function of the displacements at $t=1.5$ ps (Fig.~\ref{f.velocities}, right column), that is when the wave front is far from the excitation source. At low frequencies, the correlation function vanishes at a distance of the order of the wavelength of the main front (Fig.~\ref{f.corr}(a)): it is dominated by the low scattering of the exciting wave and the related displacements keep a trace of the incident wavelength. However, a local minimum in the correlation function can also be observed at $0.3$ nm.

The behavior at 6.5 THz is totally different: we do not see the plane wave structure even for the initial wave packet (Fig.~\ref{f.velocities}(e)). It means that the period of the external force is larger than the lifetime of plane wave excitations. It is the case of frequencies above the Ioffe-Regel criterion. In this regime, the wave packet can spread by means of diffusion only. Note in this case, that the global minimum in the displacements correlation function (Fig.~\ref{f.corr}(b)) becomes progressively pinned at the very small intrinsic length of $0.3$ nm. Indeed, the local minimum at $0.3$ nm (and its counterpart at $0.6$ nm) already present at very low frequencies (Fig.~\ref{f.corr}(a)) becomes definitely dominant at higher frequencies. This very small intrinsic length dominates the correlation function in the studied model of amorphous silicon at all frequencies above the Ioffe-Regel criterion, as already discussed in~\cite{Tanguy2015}.

\begin{figure}[t]
    \includegraphics[scale=0.55]{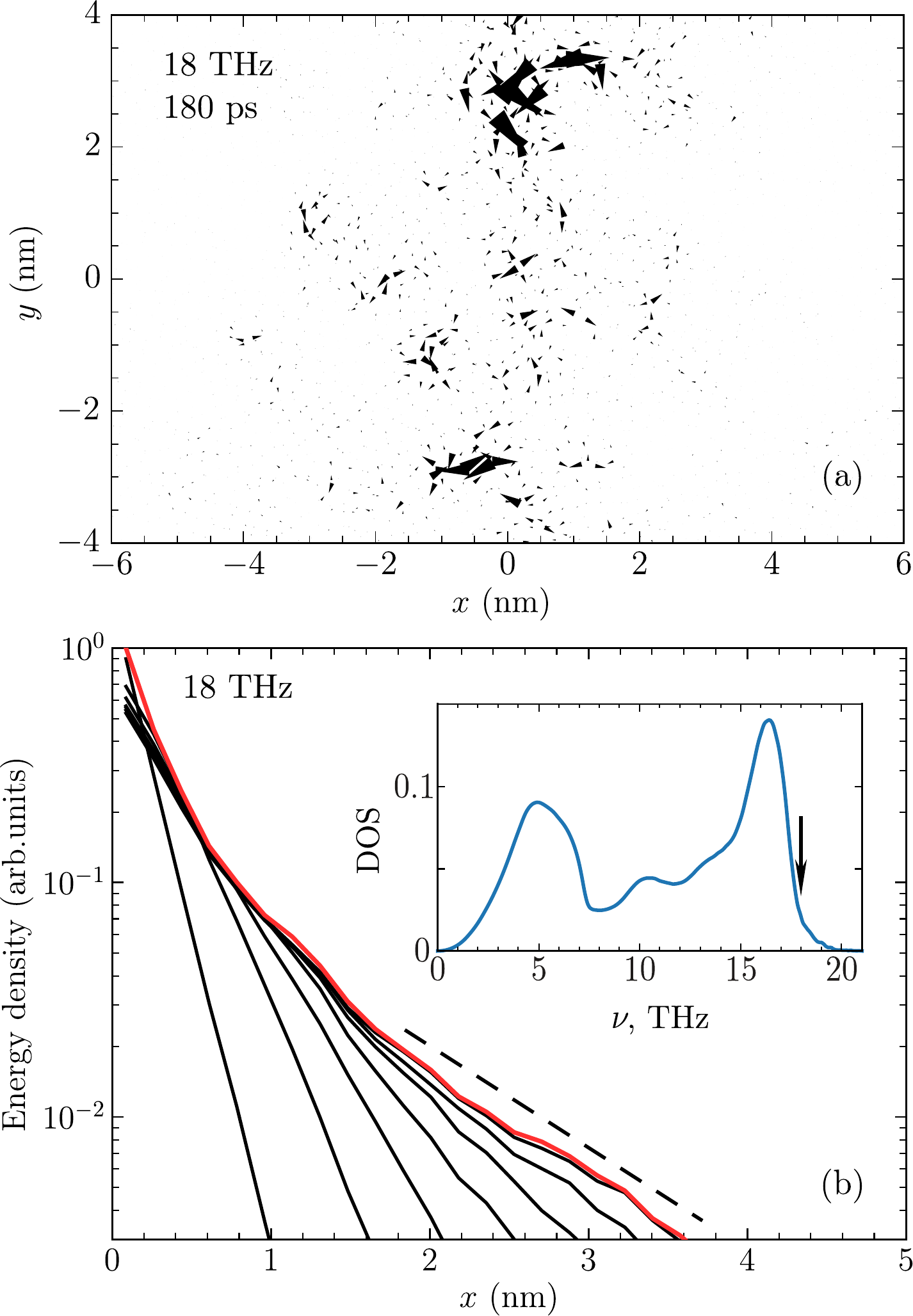}
    \caption{(Color online) Localized regime of the wave-packet propagation. Black lines show the kinetic energy density $E_T(\omega, x, t)$ for transverse wave packets at 18 THz at different times $t$: 0, 3, 12, 36, 90, 180, 360 ps. Red (gray) line shows the envelope $P_\eta(\omega, x)$. The kinetic energy density $E_T(\omega, x, t)$ is symmetric over $x$ so only $x>0$ is shown. The inset shows the vibrational density of states with 18 THz marked on it. }
    \label{f.Loc}
\end{figure}

Finally, at frequencies above the mobility edge, we have shown in a previous article~\cite{Beltukov2016} that vibrations are {\it localized}, with a multi-fractal shape. In this frequency range, we show here, that the vibrational response is very specific: it is related to a limitation of the excitations to sparsed atomic oscillations in the initally excited layer at $x \approx 0$, and the vibrations are transmitted very partly to the rest of the solid (Fig.~\ref{f.Loc}).

\section{Envelope of the Kinetic Energy}

\subsection{Attenuation of the Kinetic Energy}

\begin{figure}[t]
    \includegraphics[scale=0.55]{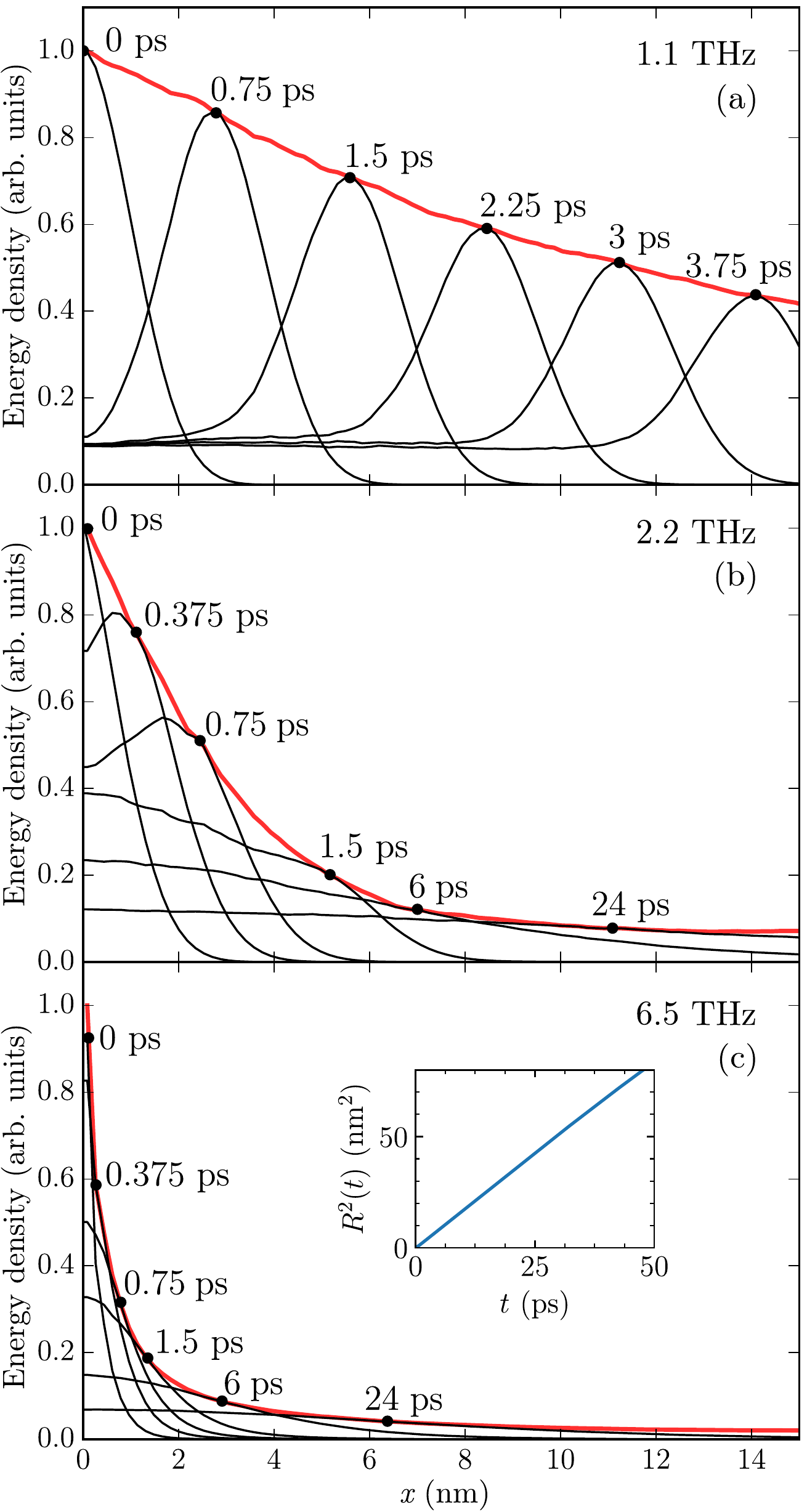}
    \caption{(Color online) Different regimes of the wave-packet propagation. Black lines show the kinetic energy density $E_T(\omega, x, t)$ for transverse wave packets at different times $t$ for three different frequencies: 1.1 THz (a), 2.2 THz (b), 6.5 THz (c). Red (gray) line shows the envelope $P_\eta(\omega, x)$. Black dots show the points, where the kinetic energy density $E_T(\omega, x, t)$ touches the envelope $P_\eta(\omega, x)$. The kinetic energy density $E_T(\omega, x, t)$ is symmetric over $x$ so only $x>0$ is shown. Inset in the panel (c) shows the wave packet squared width $R^2(t)$ for the corresponding frequency $\nu = 6.5$ THz.}
    \label{f.Propa}
\end{figure}

In order to study the transportation of vibrational energy in all the regimes previously identified, we calculate the kinetic energy density $E_\eta(\omega, x, t)$ as a function of space and time. We report in Fig.~\ref{f.Loc} the spatial dependence of the kinetic energy density at different times $t$, for a frequency above the mobility edge, $\nu=\omega/2\pi=18$~THz, and in \ref{f.Propa} for the same polarization and the same frequencies, which were used in Fig.~\ref{f.velocities}. 

Fig.~\ref{f.Loc} is a clear example of a \textit{localized} regime: if we take the positions of the maxima of the kinetic energy density at successive times, we find that it can be fitted by an exponential decay $P_0\exp(-x/\xi)$ with some localization length $\xi$. The transportation of energy is thus limited in this case to the characteristic distance $\xi$ ($\xi\approx1.0$ nm for $\nu = 18$ THz in Fig.~\ref{f.Loc}(b)). This behavior is similar to the behavior of evanescent waves~\cite{BookAcoust} and differs strongly from the other regimes.

Fig.~\ref{f.Propa}(a) is a clear example of a \textit{propagative} regime. The kinetic energy density has a maximum, whose position moves right with the sound velocity. The maximum of the kinetic energy gradually decreases with time due to the scattering by the structural disorder.

Fig.~\ref{f.Propa}(c) is a clear example of a \textit{diffusive} regime. The maximum of the kinetic energy density stays at $x=0$ whereas the width increases with time. Indeed, as shown in the inset of Fig.~\ref{f.Propa}(c), the squared width 
\begin{equation}
    R^2(t) = \frac{\int_{-\infty}^{\infty}x^2 E_\eta(\omega, x, t)\,dx}{\int_{-\infty}^{\infty}E_\eta(\omega, x, t)\,dx}
\end{equation}
is proportional to $t$. The same holds true for all frequencies up to the mobility edge (not shown here).

Fig.~\ref{f.Propa}(b) shows that, at intermediate frequencies, the propagative regime after some critical time $t = t_1$ becomes diffusive. In the case illustrated in the figure, for example, for $t<1$ ps, the position of the maximum of the kinetic energy moves to the right. However, for $t>1$ ps, the amplitude of the propagative part becomes smaller than the amplitude of the diffusive part, inducing the broadening of the kinetic energy density with a maximum located at $x=0$.

\begin{figure*}[t]
    \includegraphics[scale=0.55]{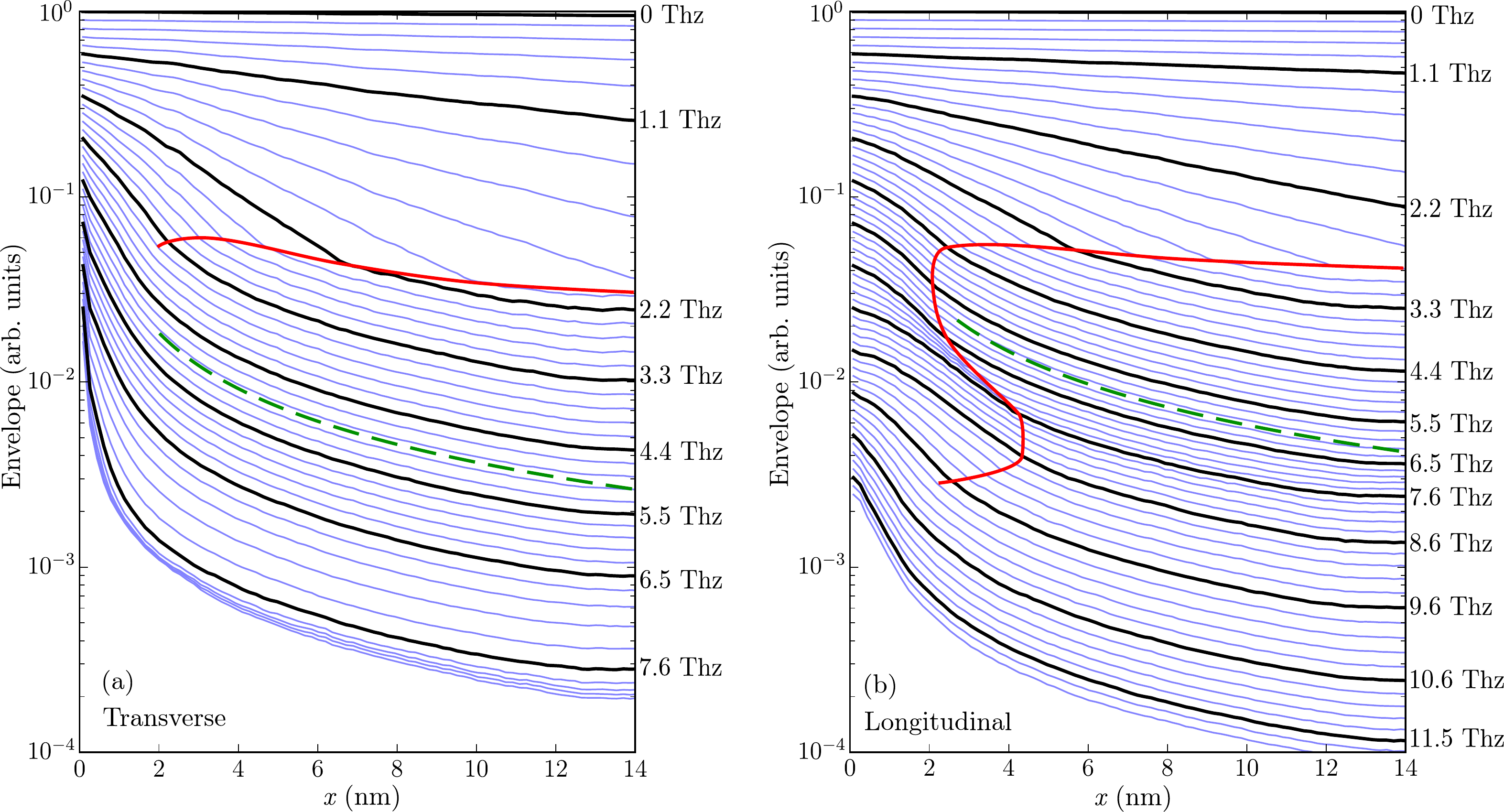}
    \caption{(Color online) Envelope of the attenuation of transverse (a) and longitudinal (b) wave packets with different frequencies. Each fifth curve is colored by black and marked by the corresponding frequency on the right. Each $n$-th curve is multiplied by $0.9^n$ to make an offset for a better visual effect. Thick red (grey) curved line shows the transition from the propagative regime to the diffusive regime. Dashed green (grey) line shows $L/x$ behaviour in the diffusive regime. }
    \label{f.envelopes}
\end{figure*}

In order to characterize quantitatively the aforementioned regimes, we calculate the \textit{envelope}
\begin{equation}
    P_\eta(\omega, x) = \max\limits_t E_\eta(\omega, x, t).
\end{equation}
as reported in Fig.~\ref{f.Propa} by a red line (gray) line. In the propagative regime, the kinetic energy density for $x > 0$ is described as
\begin{equation}
    E_{\rm prop}(x, t) = \frac{{\cal E}_0}{\sqrt{\pi v^2\tau_{\rm exc}^2} }\exp\left(-\frac{x}{l}-\frac{(x - v t)^2}{v^2\tau_{\rm exc}^2}\right).
\end{equation}
where ${\cal E}_0=2\int_0^{\infty}  E_{\rm prop}(0, t)\,v\,dt$ is the total kinetic energy delivered per unit surface to the system from the excitation force.
The envelope
\begin{equation}
    P_{\rm prop}(x) = E_{\rm prop}(x, x/v) = \frac{{\cal E}_0}{\sqrt{\pi v^2\tau_{\rm exc}^2} }e^{-x/l}   \label{eq:BL}
\end{equation}
is similar to a Beer-Lambert law with a mean free path $l$. In the diffusive regime the kinetic energy density can be approximated by the Gaussian
\begin{equation}
    E_{\rm diff}(x, t) = \frac{{\cal E}_0}{\sqrt{4\pi D t} }\exp\left(-\frac{x^2}{4 D t}\right)
\end{equation}
where the energy ${\cal E}_0$ per unit surface was the total energy already delivered to the system per unit surface (${\cal E}_0=2\int_{0}^{\infty} E_{\rm diff}(x,t)\,dx$), and $D$ is the diffusivity. The kinetic energy $E_{\rm diff}(x,t)$ per unit surface, at the position $x$ will reach its maximum value at the time $t^* = x^2/2D$. Consequently, the envelope has the form
\begin{equation}
    P_{\rm diff}(x) = E_{\rm diff}(x, t^*) = \frac{{\cal E}_0}{\sqrt{2\pi e}} \frac{1}{x}
\label{eq.diffatt}
\end{equation}

The envelopes for different frequencies $\omega$ and both polarizations $\eta$ are shown in Fig.~\ref{f.envelopes} in a logarithmic scale. In this scale, the Beer-Lambert law observed in the propagative regime is a straight line. It is worth underlying that the curves should be considered only for $x \ge x_{\rm min}$ with $x_{\rm min} \gtrsim v_\eta \tau_{\rm exc}$ bounding the region with an uncertain regime due to the finite excitation time $\tau_{\rm exc}$. The red (gray) curve indicates the transition from propagative to diffusive regime (see the next section). An envelope $\sim\!1/x$, which is typical for the diffusive regime is shown by a dashed line. One can see that there is no propagative regime at all for frequencies above $\nu_{\rm IR}^\eta$ (corresponding to 3.5~THz and 10~THz for transverse and longitudinal polarization respectively), and that there is a coexistence between propagative front and diffusive background at intermediate frequencies below $\nu_{\rm IR}^\eta$. The frequency range of the coexistence regime is especially extended for longitudinal modes. Note that for transverse waves, this regime coincides as well to the frequency interval where the Boson Peak takes places~\cite{Beltukov2016}. The Boson peak cross-over is thus compatible with the coexistence of diffusive and propagative transverse vibrations. We will now propose an analytical estimate of the {\it position} at which the diffusive part becomes dominant.

\subsection{Analysis of the transition from Propagative to Diffusive regime}

\begin{figure}[t]
    \includegraphics[scale=0.55]{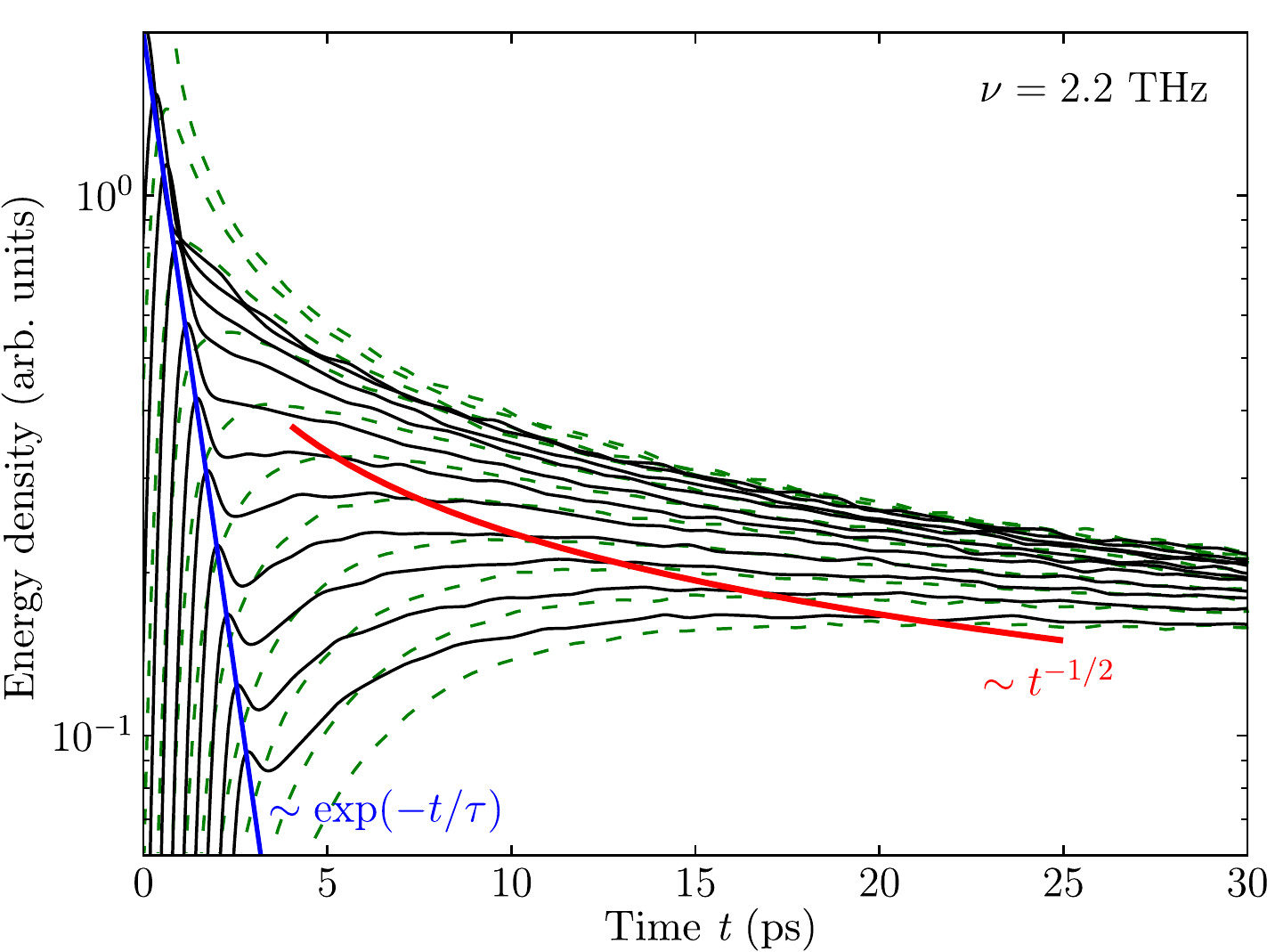}
    \caption{(Color online) Time dependence of the kinetic energy density resulting from a coherent excitation of transverse modes at 2.2 THz for different values of $x$ coordinates. From top to bottom: $x=0,1,2\ldots,10$ nm. Thick solid line shows the fit as an $\exp(-t/\tau)$ dependence (green) and a $t^{-1/2}$ dependence (red). Thin (green) dashed lines represent the time dependence of the kinetic energy density for an incoherent excitation with the same amplitude and same frequencies. }
    \label{f.time_dependence}
\end{figure}

For a given $x$ coordinate the time dependence $E_\eta(\omega, x, t)$ can reveal the coexistence of propagative and diffusive energy transfer. Figure \ref{f.time_dependence} shows such dependence for transverse modes with $\nu = 2.2$ THz. For $t < x/v_g(\omega)$ the energy density $E_\eta(\omega, x, t)$ is exponentially small because no signal can be transferred with a velocity faster than the group velocity $v_g(\omega)$. At time $t = x/v_g(\omega)$ there is a narrow peak, which is an initial (partially scattered) coherent plane wave moving at a well-defined velocity coinciding with the group velocity at that frequency. At time $t \gg x/v_g(\omega)$ one can observe a second much broader  peak of the energy density, whose position is marked by the red line (see Fig.~\ref{f.time_dependence} for $x\ge 5$ nm). This second peak is formed by the diffusive spreading of the scattered energy. From Eqs.~\ref{eq:BL} and~\ref{eq.diffatt}, one can expect $E \sim \exp(-t/\tau)$ for the first peak and $E \sim t^{-1/2}$ for the second peak. Both dependencies coincide well with the measured peak position (see Fig.~\ref{f.time_dependence}). The scattered part of the energy contributes as an after-shock. Sufficiently far from the excitation, its amplitude dominates the propagative one. We have compared it to the effect of an incoherent excitation (green dashed lines in Fig.~\ref{f.time_dependence}): the amplitudes of the scattered energy and of the {\it incoherent} energy are very similar, especially at a long time where they coincide perfectly. The only difference is the delay to reach a given position, in case of an incoherent excitation, resulting in a delayed purely diffusive motion. In the opposite case of a {\it coherent} excitation, part of the coherent energy still contributes to the after-shock: the incoherent part is not completely separated from the coherent front, and it is boosted by the propagative plane wave.

For different coordinates $x$, the intensity ratio of coherent to scattered peaks can be very different, progressively decreasing with increasing the distance from the excitation. We can thus identify a critical distance $x_t$, at which the two peaks are equivalent, meaning an equal weigth of propagative and diffusive contributions to the energy density. As a result, for $x<x_t$ the energy transfer is mostly propagative, while for $x>x_t$ it is mostly diffusive. The $x_t$ position is shown by a solid thick red (gray) line in Fig.~\ref{f.envelopes} and a dotted line in Fig.~\ref{f.LENGTHS}, marking the transition from a propagative to a diffusive regime.

For large frequencies ($\nu > 3.5$ THz for transverse modes and $\nu > 9.8$ THz for longitudinal modes) there is no distinguishable propagative peak and $x_t$ does not exist (the end of solid thick red (gray) line in Fig.~\ref{f.envelopes}). As a result, for these frequencies one can observe only the diffusive regime. Interestingly, such frequencies are close to the Ioffe-Regel limits found in \cite{Beltukov2016}.

A crude estimate of the position $x_t$ at which the transition between the dominantly propagative and the dominantly  diffusive regime occurs (thick red (gray) line in Fig.~\ref{f.envelopes}) is given by $P_{\rm prop}(x_t) = P_{\rm diff}(x_t)$. This yields
\begin{equation}
    x_t=-l\,W_{-1}\left(-\frac{v\tau_{\rm exc}}{l\sqrt{2e}}\right)
\end{equation}
where $W_{-1}$ is the lower branch of the two-valued Lambert $W$ function. Under the assumption $v \tau_{\rm exc} \ll l$ we have
\begin{equation}
    x_t \approx l \ln \frac{l}{v \tau_{\rm exc}},
\end{equation}
 which means that $x_t$ is proportional to the mean free path $l$ with a logarithmic correction.
The diffusive regime thus becomes the dominant way of energy transfer when $x \gtrsim l$. While this corresponds to the definition of the Ioffe-Regel limit, it is worth noting that the coexistence between the two regimes occurs far below the Ioffe-Regel limit, and on a large intermediate frequency band, at least for longitudinal waves. Especially, close to the Boson Peak frequency, diffusive and propagative regimes coexist for transverse as well as for longitudinal waves.

\subsection{Mean-Free paths and Penetration Depth}

\begin{figure}[t]
   \includegraphics[scale=0.55]{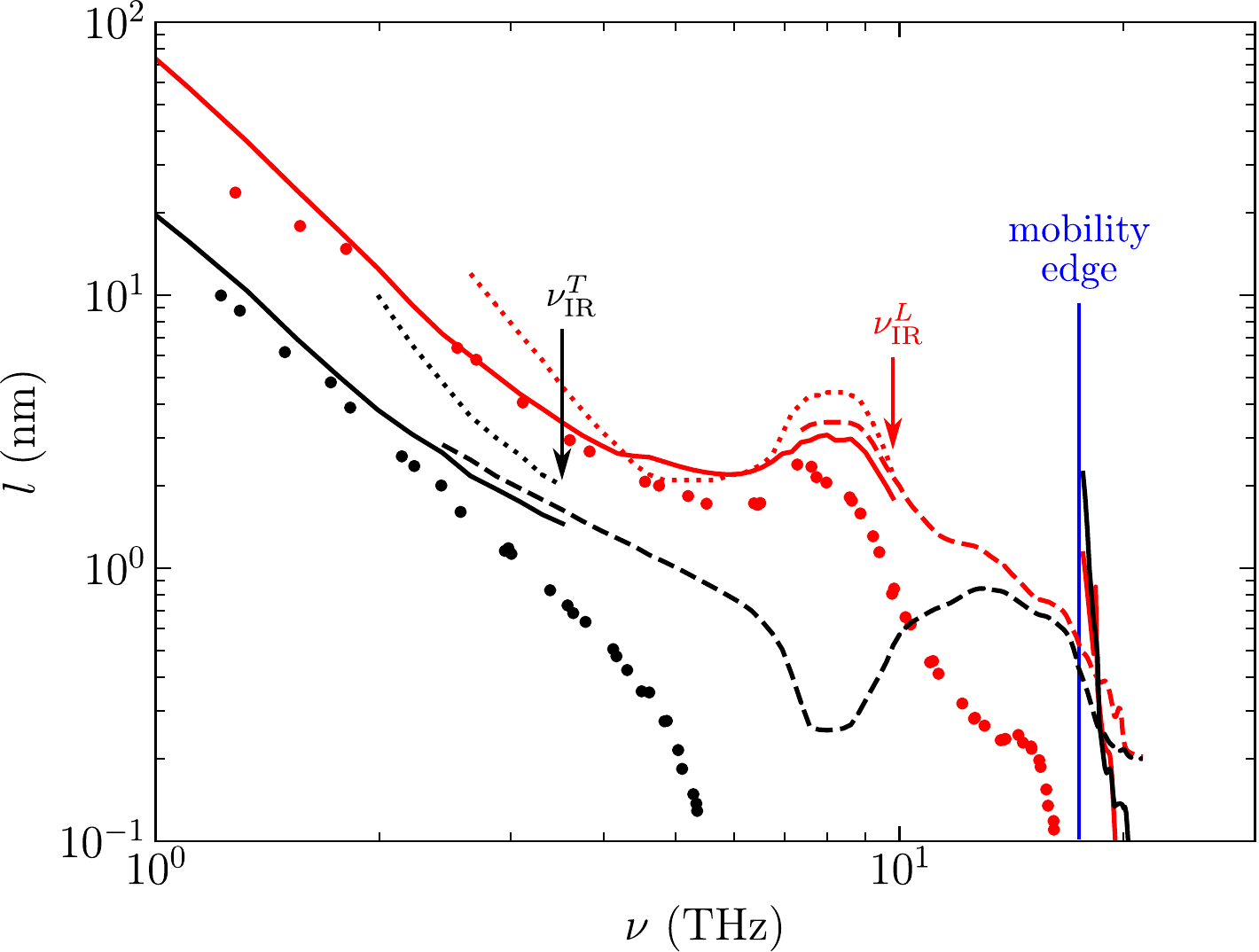}
   \caption{(Color online)  Result of the fits obtained from wave packets propagation compared to mean-free paths obtained from the analysis of $S(q,\omega)$ for transverse waves (black) and longitudinal waves (red/gray). Points show the result from the DHO fit of the Dynamical Structure Factor, $l_{\rm DSF}$. Solid lines show the mean free path $l$ obtained by the exponential fit of the envelope below Ioffe-Regel criterion and above the mobility edge. Dashed lines show the penetration length $l_{\rm pen}$. Dotted lines show the transition length $x_t$ between propagative and diffusive regime.}
    \label{f.LENGTHS}
\end{figure}

From the measurement of the envelope of the kinetic energy density, it is possible to estimate the mean-free path $l$ using the Beer-Lambert exponential fit (\ref{eq:BL}) in the propagative regime. The resulting values of $l$ as a function of frequency $\nu$ are shown in Fig.~\ref{f.LENGTHS} for transverse and longitudinal waves. The results are compared to the measurement of the mean-free path $l_{\rm DSF}$ obtained from the Dynamical Structure Factor as detailed in Ref.~\cite{Beltukov2016}. In the low-frequency range below the Ioffe-Regel criterion \AT{($\nu\ll\nu_{\rm IR}^\eta$)}, $l$ and $l_{\rm DSF}$ coincide with each other for both transverse and longitudinal waves. However, near the Ioffe-Regel criterion  \AT{($\nu\approx\nu_{\rm IR}^\eta$)} the mean free path $l$ is bigger than $l_{\rm DSF}$ predicted by the Dynamical Structure Factor analysis.

For frequencies above the Ioffe-Regel criterion, only the diffusive regime remains, so that we cannot use anymore the Beer-Lambert law (\ref{eq:BL}) for getting the mean free path. However, we can define a penetration length $l_{\rm pen}$ such that $P_\eta(\omega, l_{\rm pen}) = \frac{1}{e}P_\eta(\omega, 0)$, which  characterizes the energy transfer for any frequency independently on the transport regime. The result is shown by dashed lines in Fig.~\ref{f.LENGTHS}. Below the Ioffe-Regel criterion $l_{\rm pen}$ is close to the mean free path $l$. One can see that transverse waves have a dip in the penetration length near $\nu\approx 8$~THz. In the same frequency region, the mean free path of longitudinal waves (and the penetration length as well) shows a peak. It is approximately the same frequency region where crystalline silicon presents a gap in the transverse density of states, namely between $7.5$~THz and $13.9$~THz~\cite{Tubino-1972}, while longitudinal phonons do not have any gap. Due to the similarity of the local atomic structure of crystalline and amorphous silicon, one can deduce that the difficulty to excite transverse vibrations for $\nu\approx 8$~THz leads to a dip in the transverse penetration length, simultaneously suppressing the scattering of longitudinal waves on transverse waves.

For frequencies above the mobility edge, one can use again the exponential fit (see solid lines above the mobility edge in Fig.~\ref{f.LENGTHS}). In this case it will characterize the localization length which diverges near the mobility edge. The detailed analysis of the mobility edge and the Anderson transition is  out of the purpose of this paper and will be presented elsewhere.

\section{Discussion and Conclusion}

We have shown in this paper that the transportation of acoustic energy in an amorphous material shows well defined separate regimes: 1) a pure propagative regime with low Beer-Lambert attenuation at very low frequencies, 2) a mixed regime coupling propagative front and diffusive after-shock at intermediate frequencies (in the Boson Peak range), 3) a purely diffusive regime above the Ioffe-Regel criterion, and finally, 4) a localized regime without transportation of energy at high frequencies above the mobility edge. 
This identification has been possible by investigating the atomic response to a wave packets excitation, relating the energy transfer to the spreading of the initial excitation into coherent and incoherent vibrations: the propagative part being the coherent contribution to the transfer, and the diffusive one being the incoherent part. The transition between a dominantly propagative transfer to a dominantly diffusive one occurs at a finite distance $x_t(\omega)$ from the initial excitation, that was shown to be related to the mean-free-path $l(\omega)$, as obtained from the Beer-Lambert-like attenuation in the propagative regime. The Boson Peak frequency range is found to correspond to the coexistence between the two regimes, thus validating both scenarios for propagative and diffusive nature of the Boson Peak~\cite{Taraskin2001, Schirmacher2007, Ruffle2008, Chumakov2011}. 
The apparent attenuation of the energy maximum in the purely diffusive regime was shown to vary as $\propto 1/x$ without any intrinsic length scale (Eq.~\ref{eq.diffatt}). Finally, in the localization regime, already identified in~\cite{Beltukov2016}, our study confirms the absence of energy transfer and the transition to an acoustic insulator.
Note that the system is purely harmonic here, that is the dissipation of energy is only an effective dissipation of energy resulting from waves spreading: in the absence of a thermostat, the total energy is conserved. 

Concerning the detail of the atomic  displacements in the different regimes, we have found that at low frequency, they show a correlation length which is the wavelength of the propagative front. The diffusive after-shock itself has the same correlation length, thus confirming a low scattering process on structural disorder, as already suggested in~\cite{b.tanguy2010,Beltukov2016,Lemaitre2016}. In the purely diffusive regime, however, the correlation length appears to be controlled by the interatomic distance with a characteristic length of $3$ {\AA} close the size of the Eshelby cores in this system~\cite{Tanguy2015,Albaret2016}. 

This study thus finally sheds light onto the mechanisms of energy transfer in amorphous materials in connection to their vibrational properties, and supports the scenario of low to strong scattering of plane waves in amorphous samples, with a coexistence of diffusive and propagative energy transfer in the Boson peak frequency range. A coherent acoustic excitation will thus be converted into a completely incoherent one, comparable to heat, when the exciting frequency is  higher than the Ioffe-Regel frequency. 

Such results are of high relevance as they allow the microscopic understanding of the transport of energy -- mechanical as well as thermal, and their intertwining -- in disordered systems, answering to an ever-growing interest from the scientific community, face to the global need of identifying mechanisms for limiting heat or sound propagation.

To complete this work, it would be necessary to properly identify the effect of a thermostat on this energy conversion, in order to infer the related temperature dependence of the thermal conductivity. 
Another perspective is related to the additional role of interfaces resulting from the addition of inclusions of different stiffness inside the amorphous matrix, as proposed in many recent studies for modifying mechanical or thermal properties of the system. In this case, acoustic dynamics can become even more complex~\cite{Elford2010,Minnich2016,Tlili2017,Damart-2016}, calling thus for further deepened studies.

\begin{acknowledgments}
    This work was supported by the Institut Carnot project PREGLISS. One of the authors (Y.M.B.) acknowledges the support of the Council of the President of the Russian Federation for Support of Young Scientists and Scientific Schools (project no. SP-3299.2016.1) and the Metchnikov fellowship of the French Embassy in Russia. Two authors (D.A.P. and Y.M.B.) thanks INSA Lyon for hospitality.

\end{acknowledgments}

\appendix

\section{Decomposition of the initial impulse \label{sec:WP}}

A standard approach to calculate the spreading of the wave packet excited by the external force (\ref{eq:exc}) consists of the integration of the Newton equations (\ref{eq:Newton_exc}) for each frequency $\omega$ independently.

We can improve the performance of the integration a lot using the linearity of the Newton equations (\ref{eq:Newton_exc}). One can integrate only one system of equations with an instant impact of an external force
\begin{equation}
    \ddot{u}_{i\alpha}(t) + \sum_{j\beta} M_{i\alpha,j\beta} u_{j\beta}(t) = g_{i\alpha}^{\rm exc}(t),
\end{equation}
where
\begin{equation}
    {\bf g}_i^{\rm exc}(t) = {\bf f}_\eta \exp\left(-\frac{x_i^2}{2w^2}\right)\delta(t).
\end{equation}
Then, in order to separate frequencies, we can apply the convolution procedure
\begin{equation}
    {\bf u}_i(\omega, t) = \int_{-\infty}^\infty{\bf u}_i(t') \exp\left(i\omega (t-t')-\frac{(t-t')^2}{2\tau_{\rm exc}^2}\right)dt'.
\end{equation}
One can see that ${\bf u}_i(\omega, t)$ is the solution of the Newton equation (\ref{eq:Newton_exc}) for a given $\omega$ and the excitation force (\ref{eq:exc}). For a better numerical performance, we use a finite-window approximation
\begin{equation}
    {\bf u}_i(\omega, t) = \int_{t-\tau}^{t+\tau}{\bf u}_i(t') {\cal W}_\tau (t-t')e^{i\omega (t-t')}dt',
\end{equation}
where ${\cal W}_\tau(t')$ is a smooth even window function with the width $2\tau$ such that ${\cal W}_\tau(0) = 1$ and ${\cal W}_\tau(t') = 0$ for $|t'|\ge\tau$. We use the Jackson window function, which is close to the Gaussian with $\tau_{\rm exc}=\tau/\pi$ \cite{Weisse-2006}.

\end{document}